\documentclass{SCGE}
\usepackage
{multicol}
\usepackage{graphicx}
\begin{document}

\def\bm{\boldsymbol}

\def\dl{\displaystyle}
\def\du{\end{document}}

\Year{} %
\Month{}
\Vol{} %
\No{} %
\BeginPage{0} %
\EndPage{} %
\AuthorMark{{\rm Qu Xiaoying,} et al.}
\DOI{ } 

\title{Extending the nuclear chart by continuum: from oxygen to titanium}

\author[1,2]{X. Y. QU }{}
\author[3]{Y. CHEN }{}
\author[3]{S. Q. ZHANG}{}
\author[3]{P. W. ZHAO}{}
\author[4]{I. J. SHIN }{}
\author[5]{Y. LIM}{}
\author[4]{Y. KIM}{}
\author[3,2,6*]{J. MENG}{}

\address[{\rm1}]{College of Science, Guizhou Minzu University, Guiyang 550025,China}
\address[{\rm2}]{School of Physics and Nuclear Energy Engineering, Beihang University, Beijing 100191, China}
\address[{\rm3}]{State Key Laboratory of Nuclear Physics and Technology, School of Physics, Peking University, Beijing 100871, China}
\address[{\rm4}]{Rare Isotope Science Project, Institute for Basic Science, Daejeon 305-811, Republic of Korea}
\address[{\rm5}]{Department of Physics Education, Daegu University, Gyeongsan 712-714, Republic of Korea}
\address[{\rm6}]{Department of Physics, University of Stellenbosch, Stellenbosch, South Africa}

\maketitle \vspace{-3.5mm}{\footnotesize\begin{center}  
\end{center}}\vspace*{-5mm}

\begin{center}
\rule{16.5cm}{0.4pt}
\parbox{16.5cm}
{\begin{abstract}  Nuclear masses ranging from O to Ti isotopes are systematically investigated with relativistic
continuum Hartree-Bogoliubov (RCHB) theory, which can provide a proper treatment of pairing correlations
in the presence of the continuum. From O to Ti isotopes, there are $402$ nuclei predicted to be bound
by the density functional PC-PK1. For the $234$ nuclei with mass measured, the root mean square
(rms) deviation is $ 2.23$ MeV. It is found that the proton drip-lines predicted with various mass models are roughly the same and basically agree with the observation. The neutron drip-lines predicted, however, are quite different. Due to the continuum couplings, the neutron drip-line nuclei predicted are extended further neutron-rich than other mass models. By comparison with finite-range droplet model (FRDM), the neutron drip-line nucleus predicted by RCHB theory
has respectively $2$(O), $10$(Ne), $10$(Na), $6$(Mg), $8$(Al), $6$(Si), $8$(P), $6$(S), $14$(K), $10$(Ca), $10$(Sc), and $12$(Ti) more neutrons.
\end{abstract}}
\end{center}\vspace*{-0.6cm}
\begin{center}
\parbox{16.5cm}{\bf\jiuhao Relativistic continuum Hartree-Bogouliubov theory; Nuclear mass table; O to Ti isotopes .
}
\end{center}

\begin{center}
\parbox{16.5cm}{\PACS{\hspace*{-2mm}\rm 21.60.Jz, 21.10.Dr, 27.30.+t, 27.40.+z, 27.50.+e }
\rule{16.5cm}{0.4pt}}\end{center}
%



\wuhao\vspace*{1.5mm}
\begin{multicols}{2}
\renewcommand{\baselinestretch}{1.08} \baselineskip 12.2pt\parindent=10.8pt

In recent decades, unstable nuclear beams have extended our knowledge of nuclear physics from
the stable nuclei to the exotic nuclei far away from the stability.
The properties of these exotic nuclei, for instance, masses and decay-lives, are essential in
understanding the nucleosynthesis via rapid neutron capture ($r$-process) [1-6].
Although considerable achievements in mass measurements have been made due to the development of new experimental
techniques and facilities, most of the neutron-rich nuclei of relevance to the $r$-process are still
beyond the experimental capability in the foreseeable future and therefore, we need to rely on robust theoretical
nuclear  models.

The widely-used global nuclear mass models can be classified into the following two categories. The first consists of macroscopic-microscopic models, such as finite-range droplet model (FRDM) [7] and
Weizs\"acker-Skyrme (WS) model [8], which are proved to work pretty
well in the description of known nuclides, but its extrapolation to very
neutron-rich nuclides is generally
questionable. The second consists of microscopic mass models, for example, the Hartree-Fock-Bogoliubov (HFB)
method based on nonrelativistic density functional theory (DFT) [9-11], which
treats the macroscopic part and the microscopic corrections in a unified framework, and is believed
to have a more reliable extrapolation to the unknown regions.

Apart from the nonrelativistic DFT, the covariant density functional theory (CDFT) has attracted research focus because of the successful description of many nuclear phenomena [12-16].
It can also include the nucleon spin degree of freedom naturally and result in the nuclear spin-orbit potential automatically
with the empirical strength in a covariant way. It can reproduce
well the isotopic shifts in the Pb region [17],
and give naturally the origin of the pseudospin symmetry [18,19] as a relativistic symmetry [20-25] and the spin symmetry in the anti-nucleon spectrum [26,27].
It can include the nuclear magnetism [28], that is, a consistent description of
currents and time-odd fields, which has an important role in the nuclear magnetic moments [29-33] and nuclear rotations [34-37].
Therefore, it is natural to investigate the nuclear masses based on CDFT.

 The first CDFT mass table was reported for $2000$ even-even nuclei with $8\leq Z \leq 120$ [38],
 but without including pairing correlations. Later, by including the pairing correlations with Bardeen-Cooper-Schrieffer (BCS)
 method, the ground-state properties of $1315$ even-even nuclei with $10 \leq Z \leq 98$ were calculated [39].
 In 2005, by employing the state-dependent BCS
method with a delta pairing force, the first systematic study of the ground-state properties for over $7000$
nuclei was performed [40].

It is well known that, the pairing correlation has a critical role in open shell nuclei. In particular,
for the exotic nuclei close to the nucleon drip-lines, where the Fermi levels are very close to the continuum threshold,
pairing correlation can scatter the valence nucleons between the bound states and continuum, and therefore provide
a significant coupling between them.
As a result, some unbound nuclei predicted without pairing correlation can become bound. For example, Meng et al. [41] found that after taking into account the pairing correlation and the
contribution from the continuum, the neutron-rich nuclei $^{60-72}$Ca predicted unbound without pairing correlation are found to be bound. Therefore, the couplings between the bound states and the continuum due to the pairing correlation can strongly affect the drip-line position.

The BCS method is a popular method in dealing with pairing correlations, but not justified for
exotic nuclei as it can not include  the contribution of continuum states properly. Conversely, Bogoliubov quasiparticle transformation can provide a unified description of the mean field and pairing correlation, and include the continuum
appropriately when treated in coordinate representation.

In Refs.[15,42-45], the relativistic Hartree-Bogoliubov (RHB) theories in coordinate space have been developed for spherical nuclei. With the relativistic continuum Hartree-Bogoliubov (RCHB) theory [15,44], the first microscopic self-consistent description of halo in $^{11}$Li has been provided [42] and the giant halos in light and medium-heavy nuclei have been predicted [41,46,47]. The RCHB theory has been generalized
to treat the odd particle system [48] and combined with the Glauder model,
the charge-changing cross sections for C to F isotopes on a carbon target have been reproduced well [49].
For deformed nuclei, much effort has been made to develop a deformed RHB
theory in continuum [50-54] and an interesting shape decoupling between the core and the halo was predicted [52,54]. Later, the deformed RHB theory in continuum has been extended to incorporate the blocking effect to treat odd nucleon system [55] and the density-dependent meson-nucleon couplings [56].

As the first step to investigate the impact of the continuum for the nuclear chart, the relativistic
continuum Hartree-Bogoliubov (RCHB) theory will be used in this paper to explore the nucleon drip-lines
by assuming spherical symmetry. Because of the huge numerical computational efforts involved, we focus on
the nuclear chart ranging from O to Ti as examples.

The starting point of the CDFT is a general effective zero-range point-coupling Lagrangian density [57],
\begin{eqnarray}\label{EQ:LAG}
  {\cal L} &=& \bar\psi(i\gamma_\mu\partial^\mu-M)\psi-\frac{1}{4}F^{\mu\nu}F_{\mu\nu}-e\frac{1-\tau_3}{2}\bar\psi\gamma^\mu\psi A_\mu\nonumber\\
           & &-\frac{1}{2}\alpha_S(\bar\psi\psi)(\bar\psi\psi)-\frac{1}{2}\alpha_V(\bar\psi\gamma_\mu\psi)(\bar\psi\gamma^\mu\psi)\nonumber\\
           & &-\frac{1}{2}\alpha_{TV}(\bar\psi\vec{\tau}\gamma_\mu\psi)(\bar\psi\vec{\tau}\gamma^\mu\psi)
              -\frac{1}{2}\alpha_{TS}(\bar\psi\vec{\tau}\psi)(\bar\psi\vec{\tau}\psi)\nonumber\\
           & &-\frac{1}{3}\beta_S(\bar\psi\psi)^3\nonumber-\frac{1}{4}\gamma_S(\bar\psi\psi)^4-\frac{1}{4}
              \gamma_V[(\bar\psi\gamma_\mu\psi)(\bar\psi\gamma^\mu\psi)]^2\nonumber\\
           & &-\frac{1}{2}\delta_S\partial_\nu(\bar\psi\psi)\partial^\nu(\bar\psi\psi)
              -\frac{1}{2}\delta_V\partial_\nu(\bar\psi\gamma_\mu\psi)\partial^\nu(\bar\psi\gamma^\mu\psi)\nonumber\\
           & &-\frac{1}{2}\delta_{TV}\partial_\nu(\bar\psi\vec\tau\gamma_\mu\psi)\partial^\nu(\bar\psi\vec\tau\gamma_\mu\psi)\nonumber\\
           & &-\frac{1}{2}\delta_{TS}\partial_\nu(\bar\psi\vec\tau\psi)\partial^\nu(\bar\psi\vec\tau\psi),
\end{eqnarray}
where $M$ is the nucleon mass, and $\alpha_S,~\alpha_V, ~\alpha_{TV}, ~\alpha_{TS}, ~\beta_{S},
~\gamma_S$, $\gamma_{V}, ~\delta_{S}, ~\delta_{V}, ~\delta_{TV}, ~\delta_{TS}$ are the coupling constants.
$A_{\mu}$ and  $F_{\mu\nu}$ are respectively the four-vector potential and field strength tensor of the
electromagnetic field.

Starting from the Lagrangian density in Eq.(\ref{EQ:LAG}), one can derive the RHB equation for the
nucleons [58],
\begin{equation}
   \left(\begin{array}{cc}
           h_D-\lambda & \Delta\\
           -\Delta^*& -h^*_D+\lambda
         \end{array}\right)
   \left(\begin{array}{c}
          U_k\\
          V_k
         \end{array}\right)
   =E_k\left(\begin{array}{c}
               U_k\\
               V_k
             \end{array}\right),
\end{equation}
where $E_{k}$ is the quasiparticle energy, $\lambda$ is the Fermi level.
The Dirac hamiltonian $h_D$ is
\begin{equation}
  h_D=\bm{\alpha}\cdot\bm{p}+ \mathbf{\beta}(M+S(\bm{r})) + V (\bm{r}),
\end{equation}
where the scalar and vector potentials are, respectively,
\begin{subequations}
\begin{eqnarray}
  S(\bm{r})&=&\alpha_S \rho_S + \beta_S \rho_S^2 +\gamma_S\rho_S^3 +\delta_S\triangle \rho_S,\\
  V(\bm{r})&=&\alpha_V \rho_V + \gamma_V\rho_V^3 +\delta_V \triangle\rho_V+e A_0\nonumber\\
           & &+\alpha_{TV}\tau_3 \rho_{TV}+\delta_{TV} \tau_3 \triangle \rho_{TV}
\end{eqnarray}
\end{subequations}
with the local densities
\begin{subequations}
  \begin{eqnarray}
      \rho_S(\bm{r})     &=&\sum_{k>0 }\bar V_k(\bm{r})V_k(\bm{r}),\\
      \rho_{V}(\bm{r})   &=&\sum_{k>0 } V_k^{\dagger}(\bm{r})V_k(\bm{r}),\\
      \rho_{T V}(\bm{r}) &=&\sum_{k>0 } V_k^{\dagger}(\bm{r})\tau_3 V_k(\bm{r}).
  \end{eqnarray}
\end{subequations}

The pairing potential reads,
\begin{eqnarray}
\Delta_{kk'}(\bm{r},\bm{r'})
&=&-\sum_{\tilde{k}\tilde{k'}}\mathbf{V}_{kk',\tilde{k}\tilde{k'}}(\bm{r},\bm{r'})\kappa_{\tilde{k}\tilde{k'}}(\mathbf{r},\mathbf{r'})
\end{eqnarray}\label{eq3}
with the pairing tensor $ \kappa=U^{*}V^T$ and a density-dependent delta pairing force
\begin{eqnarray}
    V^{pp}(\mathbf{r_1},\mathbf{r_2})&=&V_0\delta(\mathbf{r_1}-\mathbf{r_2})\frac{1}{4}(1-P^{\sigma})
                                        (1-\frac{\rho(\mathbf{r_1})}{\rho_0}).
\end{eqnarray}
The RCHB theory [42,44]
solves the RHB equations in coordinate representation, thus it provides a fully self-consistent description of
both the continuum and the bound states as well as the coupling between them.

\begin{figure*}
 \includegraphics[width=20cm]{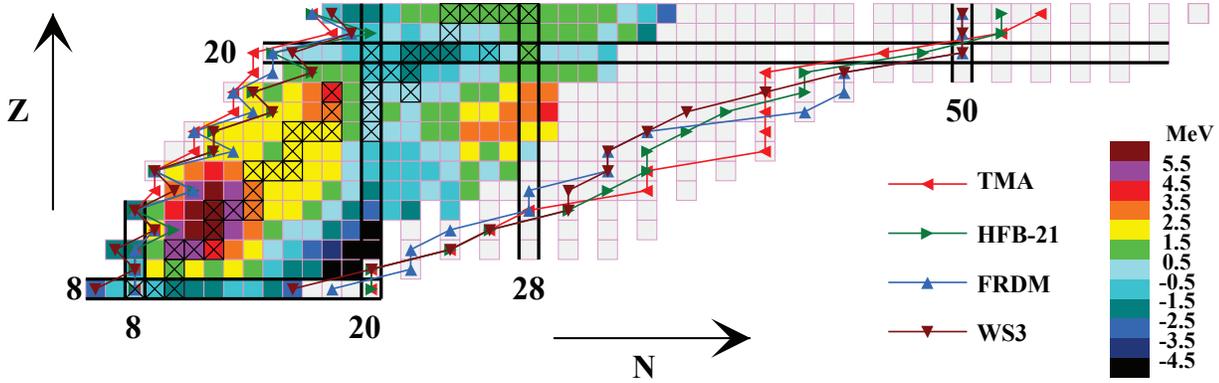}
 \caption{ $402$ nuclei from O to Ti predicted to be bound by the RCHB theory
           with the covariant density functional PC-PK1.
           For $ 234$ nuclei with the available data, the binding energy differences $E_{\rm{b}}^{\rm{Exp.}}-E_{\rm{b}}^{\rm{Cal.}}$
           between the data [62] and present calculation are shown as different color. Furthermore,
           the nucleon drip-lines predicted by the mass tables TMA, HFB-21, FRDM and WS3 are plotted for comparison.}\label{fig1}
\end{figure*}

Following the procedures as described in Ref. [44], the RCHB equations are solved in a box with a
size of $R=20$ fm and a step size of $0.1$ fm. In the present work, we use the density functional PC-PK1 [57] for particle-hole
channel, which particularly improves the description for
isospin dependence of the nuclear masses [59] and has been successfully used in describing the Coulomb displacement energies between mirror nuclei [60],
fission barriers [61] as well as nuclear rotations [35-37]. For particle-particle
channel, the density-dependent delta pairing force with the saturation density $\rho_0=0.152$ fm$^{-3}$ is used and
the strength $V_0=685.0$~MeV~fm$^{3}$ is fixed by reproducing experimental odd-even mass differences of Ca isotopes
obtained with a three-point formula. The contribution from the continuum is restricted within a cutoff energy
$E_{\rm{cut}}\sim 100$ MeV.

The particular purpose of the present work is to investigate the extension of the nuclear chart by the continuum couplings.
Therefore, we focus on the location of the nucleon drip-lines. On one hand, both the one-nucleon separation energy
\begin{subequations}
 \begin{eqnarray}
   S_n (Z,N) &=& B (Z,N)-B(Z,N-1),\\
   S_p (Z,N) &=& B(Z,N)-B(Z-1,N),
 \end{eqnarray}
\end{subequations}
and two-nucleon separation energy
\begin{subequations}
 \begin{eqnarray}
   S_{2n}(Z,N)&=& B(Z,N)-B(Z,N-2),\\
   S_{2p}(Z,N)&=&B(Z,N)-B(Z-2,N),
 \end{eqnarray}
\end{subequations}
can provide the information on nucleon drip-lines.
On the other hand, the Fermi level, $\lambda_n$ and $\lambda_p$, can provide the bound information of the nucleus. In the present work, only if the nucleon separation energy
is positive and the Fermi level is negative, the nucleus is considered to be bound.

In Fig.~\ref{fig1}, the bound nuclei region from O to Ti isotopes predicted by RCHB theory with the density functional PC-PK1 are shown as squares. It is found that $402$ bound nuclei
are predicted. The squares with cross represent the $40$ stable nuclei existing in nature. If we regard the most stable nucleus in each isotope chain, that is, the nucleus with the maximum binding energy per
nucleon, as a benchmark, there are $100$ nuclei in the neutron-deficient side and $287$ nuclei in the neutron-rich side.
Among these nuclei, the masses of $234$ nuclei are already known experimentally [62],
while those of $7$ neutron-deficient and $161$ neutron-rich nuclei are not measured yet.
It should be mentioned that many nuclei observed experimentally are reported [63], including $5$ neutron-deficient nuclei, that is, $^{23}$Si, $^{26}$P, $^{27}$S, $^{31}$Ar, $^{35}$Ca, and $53$ neutron-rich nuclei, that is, $^{29,31}$F, $^{32,34}$Ne,
$^{33,35,37}$Na, $^{38,40}$Mg, $^{39,40}$Al, $^{42,43}$Si, $^{44-46}$P, $^{46-48}$S, $^{47-51}$Cl, $^{48-53}$Ar,
$^{52-56}$K, $^{53-58}$Ca, $^{56-61}$Sc, $^{58-63}$Ti. Their mass measurements are expected to be achieved in the near future. Furthermore, it can be noted that $ 13$ exotic
nuclei proved to be bound in experiments, that is, $^{12,13}$O, $^{14-16}$F, $^{16,31}$Ne, $^{18,19,32}$Na, $^{19,35}$Mg,
$^{39}$Sc, are absent in the present prediction, which needs to be further examined in the future, for instance, by taking
into account the deformation effect.

In order to evaluate the agreement of the present calculated masses with the available data,
we show the binding energy differences $E_{\rm{b}}^{\rm{Exp.}}-E_{\rm{b}}^{\rm{Cal.}}$ with different
colors in Fig.~\ref{fig1}. One can find that most deviations are in the range of $-2.5\sim 2.5$ MeV,
resulting in the rms deviation $\sigma$ in this nuclear region is $2.23$ MeV.
There are two nuclear regions with the deviation $|E_{\rm{b}}^{\rm{Exp.}}-E_{\rm{b}}^{\rm{Cal.}}|>3.5$ MeV.
One is around the $N=Z=12$ region, and the other is near $N=20$ and $Z=10$ region. The reasons for such
large deviation areas may result from the following reasons: firstly, the proton-neutron pairing correlation,
which can provide additional binding in $N \simeq Z$ nuclei [64], is not taken into
account; secondly, the deformation effect is not included. Indeed, when we
restrict ourselves to the spherical nuclei, for instance, O and Ca isotopes, the results achieve good agreement
with the data and the corresponding rms deviation $\sigma$ reduces to $1.67$ and $1.09$ MeV, respectively.

Following the definition of the nucleon drip-line, in Fig.~\ref{fig1}, one can easily recognize the
proton and neutron drip-lines for O to Ti isotope chains predicted with the RCHB theory. For comparison, we
also show the predictions from the mass tables FRDM [7], WS3 [65],
HFB-21 [11] and TMA [40] in Fig.~\ref{fig1}. On the
neutron-deficient side, owing to the repulsive electrostatic interaction between the protons,
the proton drip-line lies relatively close to the valley of stability. The predictions from different
theoretical models are roughly the same as the experimental observations.

On the neutron-rich side, as neutrons do not carry the electric charge, the neutron drip-line
is located far from the valley of stability. Since pairing correlation tends to make the even-$N$
isotopes more bound than their neighbouring odd-$N$ isotopes, several odd-$N$ isotopes
are found to be absent before reaching the neutron drip-line. Furthermore, it is
found that the bound nuclei predicted theoretically are much more than the present experimental observation.
The neutron drip-lines predicted by RCHB theory are roughly the same as other mass models
for O and F but are more neutron-rich for the other nuclei. By comparison with FRDM, the neutron drip-line nucleus predicted by RCHB theory has respectively $2$(O), $10$(Ne), $10$(Na), $6$(Mg), $8$(Al), $6$(Si), $8$(P), $6$(S), $14$(K), $10$(Ca), $10$(Sc), and $12$(Ti) neutrons more. This extension of the nuclear chart predicted by the RCHB theory is because of the proper treatment of the continuum couplings and exhibits the hint of the appearance of giant halos in light neutron-rich
nuclei [41,47].

In conclusion, the nuclear masses are systematically investigated in nuclear region from O to Ti isotopes with
 relativistic continuum Hartree-Bogoliubov (RCHB) theory, which can provide an appropriate
treatment of pairing correlations in the presence of the continuum. By applying the density functional PC-PK1,
$402$ bound nuclei are predicted in this nuclear area, including $100$ neutron-deficient and $287$ neutron-rich bound nuclei. For the $234$ nuclei with mass measured, the RCHB results
can reproduce the data with the rms deviation $2.23$ MeV. For the spherical
O and Ca isotopes, the results can achieve good agreement with the data and the corresponding rms deviation
is reduced to $1.67$ and $1.09$ MeV, respectively. It is found that the drip-lines predicted by various mass models
are roughly the same as the experimental observation on neutron-deficient side. For the neutron-rich side, because of the
continuum couplings, the neutron drip-line predicted by RCHB theory is extended further neutron-rich than other
mass models. By comparison with FRDM, the neutron drip-line nucleus predicted by RCHB theory
has respectively $2$(O), $10$(Ne), $10$(Na), $6$(Mg), $8$(Al), $6$(Si), $8$(P), $6$(S), $14$(K), $10$(Ca), $10$(Sc), and $12$(Ti) neutrons more.

\Acknowledgements{\bahao
This work was partially supported by the Major State 973 Program 2007CB815000, National
Natural Science Foundation of China under Grant Nos. 11105005, 11105006, 11175002,
China Postdoctoral Science Foundation under Grant Nos. 20100480149, 201104031, and the Research
Fund for the Doctoral Program of Higher Education under Grant No. 20110001110087.
X. Y. Qu would like to acknowledge the discussions with Z. X. Li, L. S. Song, and Z. M. Niu,
as well as the support of the Young Core Instructor and Domestic Visitor Foundation
from the Wuhan center of Teacher Education Exchange.
The work of Y. Kim and I. J. Shin was supported by the Rare Isotope Science Project funded by the
Ministry of Science, ICT and Future Planning (MSIP) and National Research Foundation
(NRF) of KOREA. Y. Lim is supported by the Basic Science Research Program through the National Research Foundation of Korea (NRF) funded by the Ministry of Education, Science and Technology Grant No. 2010-0023661.}


\normalsize \vskip0.3in\parskip=0mm \baselineskip 18pt
\renewcommand{\baselinestretch}{1.1}\footnotesize\parindent=4mm\bahao


\REF{1\ }Kratz K.-L., Bitouzet J.-P., Thielemann F.-K., M\"oller P., Pfeiffer B., Isotopic r-process abundances and nuclear structure far from stability - Implications for the r-process mechanism. Astrophys. J., 1993, 403: 216-238
\REF{2\ }Arnould M., Goriely S., Takahashi K., The r-process of stellar nucleosynthesis: Astrophysics and nuclear physics achievements and mysteries. Phys. Rep., 2007, 450: 97--213
\REF{3\ }Li Z., Niu Z.~M., Sun B., Wang N., Meng J., WLW mass model in nuclear r-process calculations. Acta Physica Sinica, 2012, 61: 72601
\REF{4\ }Sun B., Meng J., Challenge on the Astrophysical R-Process Calculation with Nuclear Mass Models. Chin. Phys. Lett., 2008, 25: 2429--2431
\REF{5\ }Sun B., Montes F., Geng L.~S., Geissel H., Litvinov Y.~A., Meng J., Application of the relativistic mean-field mass model to the $r$-process and the influence of mass uncertainties. Phys. Rev. C, 2008, 78: 025806
\REF{6\ }Niu Z.~M., Niu Y.~F., Liang H. Z., Long W. H., Nik\v{s}i\'{c} T., Vretenar D., Meng J., $\beta$-decay half-lives of neutron-rich nuclei and matter flow in the r-process. Phys. Lett. B, 2013,723: 172
\REF{7\ }M\"oller P.,Nix J.~R., Myers W.~D., Swiatecki W.~J., Nuclear Ground-State Masses and Deformations. At. Data Nucl. Data Tables, 1995, 59: 185--381
\REF{8\ }Wang N., Liang H. Z., Liu M., Wu X., Mirror nuclei constraint in nuclear mass formula. Phys. Rev. C, 2012, 82: 044304
\REF{9\ }Goriely S., Chamel N., Pearson J.~M.,Skyrme-Hartree-Fock-Bogoliubov Nuclear Mass Formulas: Crossing the 0.6~MeV Accuracy Threshold with Microscopically Deduced Pairing. Phys. Rev. Lett., 2009, 102: 152503
\REF{10\ }Goriely S., Hilaire S., Girod M., P\'eru S., First Gogny-Hartree-Fock-Bogoliubov Nuclear Mass Model. Phys. Rev. Lett., 2009, 102: 242501
\REF{11\ }Goriely S., Chamel N., Pearson J.~M., Further explorations of Skyrme-Hartree-Fock-Bogoliubov mass formulas. XII. Stiffness and stability of neutron-star matter. Phys. Rev. C, 2010, 82: 035804
\REF{12\ }Serot B., Walecka J., The relativistic nuclear many-body problem. Adv. Nucl. Phys., 1986, 16: 1
\REF{13\ }Ring P., Relativistic mean field theory in finite nuclei. Prog. Part. Nucl. Phys., 1996, 37: 193--263
\REF{14\ }Vretenar D., Afanasjev A.~V., Lalazissis G.~A., Ring P., Relativistic Hartree-Bogoliubov theory: static and dynamic aspects of exotic nuclear structure. Phys. Rep., 2005, 409: 101--259
\REF{15\ }Meng J., Toki H., Zhou S.~G., Zhang S.~Q., Long W.~H., Geng L.~S., Relativistic continuum Hartree Bogoliubov theory for ground-state properties of exotic nuclei. Prog. Part. Nucl. Phys., 2006, 57: 470--563
\REF{16\ }Nik\v{s}i\'{c} T., Vretenar D., Ring P., Relativistic nuclear energy density functionals: Mean-field and beyond. Prog. Part. Nucl. Phys., 2011, 66: 519 --548
\REF{17\ }Sharma M.~M., Lalazissis G.~A., Ring p., Anomaly in the charge radii of Pb isotopes. Phys. Lett. B, 1993, 317: 9--13
\REF{18\ }Arima A., Harvey M., ShimizuK., Pseudo LS coupling and pseudo SU3 coupling schemes. Phys. Lett. B, 1969, 30: 517--522
\REF{19\ }Hecht K.~T., Adler A., Generalized seniority for favored $J \neq 0$ pairs in mixed configurations. Nucl. Phys. A, 1969, 137: 129--143
\REF{20\ }Ginocchio J., Pseudospin as a Relativistic Symmetry. Phys. Rev. Lett., 1997, 78: 436--439
\REF{21\ }Meng J., Sugawara-Tanabe K., Yamaji S., Ring P., Arima A., Pseudospin symmetry in relativistic mean field theory. Phys. Rev. C, 1998, 58: 628--631
\REF{22\ }Meng J., Sugawara-Tanabe K., Yamaji S., Arima A., Pseudospin symmetry in Zr and Sn isotopes from the proton drip line to the neutron drip line. Phys. Rev. C, 1999, 59: 154--163
\REF{23\ }Long W.~H., Sagawa H., Meng J., Van~Giai N., Pseudo-spin symmetry in density-dependent relativistic Hartree-Fock theory. Phys. Lett. B, 2006, 639: 242 -- 247
\REF{24\ }Liang H.~Z., Zhao P.~W., Zhang Y., Meng J. , Van~Giai N., Perturbative interpretation of relativistic symmetries in nuclei. Phys. Rev. C, 2011, 83: 041301
\REF{25\ }Lu B.-N., Zhao E.-G., Zhou S.-G., Pseudospin Symmetry in Single Particle Resonant States. Phys. Rev. Lett., 2012, 109: 072501
\REF{26\ }Zhou S.-G., Meng J.,  Ring P., Spin Symmetry in the Antinucleon Spectrum. Phys. Rev. Lett., 2003, 91: 262501
\REF{27\ }Liang H.~Z., Long W.~H., Meng J., Van~Giai N., Spin symmetry in Dirac negative-energy spectrum in density-dependent relativistic Hartree-Fock theory. Eur. Phys. J. A, 2010, 44: 119--124
\REF{28\ }Koepf W., Ring P., A relativistic description of rotating nuclei: The yrast line of 20Ne. Nucl. Phys. A, 1989, 493: 61--82
\REF{29\ }Yao J.~M., Chen H., Meng J., Time-odd triaxial relativistic mean field approach for nuclear magnetic moments. Phys. Rev. C, 2006, 74: 024307
\REF{30\ }Arima A., A short history of nuclear magnetic moments and GT transitions. Sci. China-Phys. Mech. Astron., 2011, 54: 188
\REF{31\ }Li J., Meng J., Ring P., Yao J.~M., Arima A., Relativistic description of second-order correction to nuclear magnetic moments with point-coupling residual interaction. Sci. China-Phys. Mech. Astron., 2011, 54: 204
\REF{32\ }Li J., Yao J.~M., Meng J., Arima A., One-Pion Exchange Current Corrections for Nuclear Magnetic Moments in Relativistic Mean Field Theory. Prog. Theor. Phys., 2011, 125: 1185--1192
\REF{33\ }Wei J.~X., Li J.,  Meng J., Relativistic Descriptions of Nuclear Magnetic Moments. Prog. Theor. Phys., 2012, 196: 400--406
\REF{34\ }Afanasjev A., Ring P., K\"onig J., Cranked relativistic Hartree-Bogoliubov theory: formalism and application to the superdeformed bands in the A~90 region. Nucl. Phys. A, 2000, 676: 196--244
\REF{35\ }Zhao P.~W., Peng J., Liang H.~Z., Ring P., Meng J., Covariant density functional theory for antimagnetic rotation. Phys. Rev. C, 2012, 85: 054310
\REF{36\ }Zhao P.~W., Peng J., Liang H.~Z., Ring P., Meng J., Antimagnetic Rotation Band in Nuclei: A Microscopic Description. Phys. Rev. Lett., 2011, 107: 122501
\REF{37\ } Zhao P.~W., Zhang S.~Q., Peng J., Liang H.~Z., Ring P., Meng J., Novel structure for magnetic rotation bands in $^{60}$Ni. Phys. Lett. B, 2011, 699: 181--186
\REF{38\ }Hirata D. , Sumiyoshi K., Tanihata I., Sugahara Y., Tachibana T., Toki H., A systematic study of even-even nuclei up to the drip lines within the relativistic mean field framework. Nucl. Phys. A, 1997, 616: 438--445
\REF{39\ }Lalazissis G., Raman S.,  Ring P., Ground-state properties of even-even nuclei in the relativistic mean-field theory. At. Data and Nucl. Data Tables, 1999, 71: 1--40
\REF{40\ }Geng L.~S., Toki H., Meng J., Masses, Deformations and Charge Radii-Nuclear Ground-State Properties in the Relativistic Mean Field Model. Prog. Theor. Phys., 2005, 113: 785--800
\REF{41\ }Meng J., Toki H., Zeng J.~Y., Zhang S.~Q., Zhou S.-G., {G}iant halo at the neutron drip line in {C}a isotopes in relativistic continuum {H}artree-{B}ogoliubov theory. Phys. Rev. C, 2002, 65, 041302
\REF{42\ }Meng J., Ring P., Relativistic Hartree-Bogoliubov Description of the Neutron Halo in $^{11}$Li. Phys. Rev. Lett., 1996, 77: 3963--3966
\REF{43\ }P\"oschl W., Vretenar D., Lalazissis G.~A., Ring P., Relativistic Hartree-Bogoliubov Theory with Finite Range Pairing Forces in Coordinate Space: Neutron Halo in Light Nuclei. Phys. Rev. Lett., 1997, 79: 3841--3844
\REF{44\ }Meng J., Relativistic continuum Hartree-Bogoliubov theory with both zero range and finite range Gogny force and their application.  Nucl. Phys. A, 1998, 635: 3--42
\REF{45\ }Lalazissis G., Vretenar D., P\"oschl W., Ring P., Reduction of the spin-orbit potential in light drip-line nuclei. Phys. Lett. B, 1998, 418: 7--12
\REF{46\ }Meng J., Ring P., Giant Halo at the Neutron Drip Line. Phys. Rev. Lett., 1998, 80: 460--463
\REF{47\ }Zhang S.~Q., Meng J., Zhou S.-G., Proton magic even-even isotopes and giant halos of Ca isotopes with relativistic continuum Hartree-Bogoliubov theory. Sci. China G, 2003, 46: 632-658
\REF{48\ } Meng J., Tanihata I., Yamaji S., The proton and neutron distributions in Na isotopes: the development of halo and shell structure. Phys. Lett. B, 1998, 419: 1--6
\REF{49\ }Meng J., Zhou S.-G.,  Tanihata I., The relativistic continuum Hartree¨CBogoliubov description of charge-changing cross section for C, N, O and F isotopes. Phys. Lett. B, 2002, 532: 209--214
\REF{50\ }Meng J., L\"u H.~F., Zhang S.~Q., Zhou S.-G., Giant, hyperon, and deformed halos near the particle drip line. Nucl. Phys. A, 2003, 722: 366c--371c
\REF{51\ }Zhou S.-G., Meng J., Ring P., in Physics of Unstable Nuclei, Proceedings of the International Symposium on Physics
of Unstable Nuclei (ISPUN07), 2008: 402¨C408.
\REF{52\ }Zhou S.-G., Meng J., Ring P., Zhao E.-G., Neutron halo in deformed nuclei. Phys. Rev. C, 2010, 82: 011301
\REF{53\ }Zhou S.-G., Meng J., Ring P., Zhao E.-G., Neutron halo in deformed nuclei from a relativistic Hartree-Bogoliubov model in a Woods-Saxon basis. J. Phys. Conf. Ser., 2011, 312: 092067
\REF{54\ }Li L.~L., Meng J., Ring P., Zhao E.-G., Zhou S.-G., Deformed relativistic Hartree-Bogoliubov theory in continuum. Phys. Rev. C, 2012, 85: 024312
\REF{55\ }Li L.~L., Meng J., Ring P., Zhao E.-G., Zhou S.-G., Odd Systems in Deformed Relativistic Hartree Bogoliubov Theory in Continuum. Chin. Phys. Lett., 2012, 29: 42101
\REF{56\ }Chen Y., Li L.~L., Liang H.~Z., Meng J., Density-dependent deformed relativistic Hartree-Bogoliubov theory in continuum. Phys. Rev. C, 2012, 85: 067301
\REF{57\ }Zhao P.~W., Li Z.~P., Yao J.~M., Meng J., New parametrization for the nuclear covariant energy density functional with a point-coupling interaction. Phys. Rev. C, 2010, 82: 054319
\REF{58\ }Kucharek H., Ring P., Relativistic field theory of superfluidity in nuclei. Z. Phys. A, 1991, 339: 23--25
\REF{59\ }Zhao P.~W., Song L.~S., Sun B., Geissel H., Meng J., Crucial test for covariant density functional theory with new and accurate mass measurements from Sn to Pa. Phys. Rev. C, 2012, 86: 064324
\REF{60\ }Sun B., Zhao P.~W., Meng J., Mass prediction of proton-rich nuclides with the Coulomb
displacement energies in the relativistic point-coupling model. Sci. China Phys. Mech. Astron., 2011, 54: 210
\REF{61\ }Lu B.-N., Zhao E.-G., Zhou S.-G., Potential energy surfaces of actinide nuclei from a multidimensional constrained covariant density functional theory: Barrier heights and saddle point shapes. Phys. Rev. C, 2012, 85: 011301
\REF{62\ } Wang M., Audi G., Wapstra A., Kondev F., MacCormick M., Xu X., Pfeiffer B., The Ame2012 atomic mass evaluation. Chin. Phys. C, 2012, 36: 1603
\REF{63\ }Thoennessen M., Current status and future potential of nuclide discoveries. Rep. Prog. Phys., 2013, 76: 056301
\REF{64\ } Wigner E., On the Consequences of the Symmetry of the Nuclear Hamiltonian on the Spectroscopy of Nuclei. Phys. Rev., 1937, 51: 106--119
\REF{65\ } Liu M., Wang N., Deng Y., Wu X., Further improvements on a global nuclear mass model. Phys. Rev. C, 2011, 84: 014333

\end{multicols}

\end{document}